# Electronic Structure Topology Associated Domain is Useful to Minimize the Uncertainty of QM/MM Boundary Charge Transfer Effects

Jiajun Yang[1], Fang Liu[1], Dongju Zhang[2], Likai Du[1]*

[1]Hubei Key Laboratory of Agricultural Bioinformatics, College of Informatics, Huazhong Agricultural University, Wuhan, 430070, P. R. China

[2]Institute of Theoretical Chemistry, Shandong University, Jinan, 250100, P. R. China

*To whom correspondence should be addressed.

Likai Du: dulikai@mail.hzau.edu.cn

**Abstract**

   The charge transfer effect is an important component in the physical description of realistic proteins. In the hybrid quantum mechanical-molecular mechanical (QM/MM) simulations, the significant charge transfer between the QM/MM boundaries could lead to the slow convergence problem with very large QM regions. In this work, we will discuss how community structure in complex network (electronic structure topology associated domain) can be used to measure QM/MM boundary effects and how it can provide a different perspective on well established concepts in available QM/MM simulations. The graph theory is employed to provide an useful solution to distinguish the significant charge transfer regions between the core active site and the surrounding protein environment in the QM/MM simulations. According to community detection algorithm for complex network, the charge transfer topology associated domain (ctTAD) is suggested to provide an alternative tool for systematically minimizing the charge transfer effects of QM/MM boundary with minor computational costs.

**Introduction**

In QM/MM simulations, the region of chemical interest is treated by quantum mechanics (QM), while the surrounding enzyme is described by an empirical molecular mechanics (MM) model.[1-9] To provide a sufficient balance between computational resource and accuracy, the sizes of typical QM regions are usually on the order of tens of atoms, including the ligands and a few direct residues.[8, 10-11] However, recent studies have revealed an unexpected slow convergence of the QM regions for the QM/MM simulations[12-21]. And most molecular properties, such as the reaction barriers[16-17, 20] and excitation energies[15], are predicted to reach their asymptotic limits with 500 or more atoms in the QM regions. The convergence of partial charges[18, 20] and NMR shieldings[14, 19] is even slower, typically on the order of 1000 atoms.

Nowadays, the QM regions in QM/MM methodology are still largely determined by iterative calculations to estimate the importance of each residue.[16, 21-26] Some researchers have proposed to evaluate the importance of each amino acid in the QM regions by the free energy perturbation analysis [27] and charge deletion analysis [16, 28], for which the effects on relative smaller QM regions are determined by changing a fraction of the MM environment. Note that, the important residues may be properly described by MM point charges without incorporating them into the QM region, if they do not impact electronic properties. Kulik et. al. suggested to evaluate the charge density redistribution of residues by the charge shift analysis (CDA) and Fukui shift analysis (FSA) scheme, which may present the recent efforts for systematic QM region determination.[26]

The issue of slow convergence of QM regions in QM/MM simulations is reported to be partially correlated with the significant charge transfer effects between amino acids in proteins.[20, 26, 29] The charge transfer takes place in a wide range of biological processes, including photosynthesis, respiration, and signal transduction of biology, enzymatic reactions, gene replication and mutation and *so on*.[30-35] In biological systems, charge transfer could occur between donors and acceptors separated by a long distance, for example across protein-protein complexes. [36-42]

The charge transfer in proteins presents an intriguing and challenging problem to minimize QM/MM boundary effects. And the charge transfer is not a localized effect, which is at least a two-body interaction between amino acids in proteins. Thus, we believe it is necessary to focus on the overall topology of charge transfer effects in realistic proteins. In this regard, there are also many works discussing how to represent the electron transfer pathways connecting electron donating and accepting fragments in biological [43-49] and disordered material systems[50-55]. The hybrid QM/MM calculations are also applied to map the electron transfer pathways.[56-57]

The complex network analysis is an appeal and popularity to obtain qualitative insights, which has been widely used in various fields of chemical and biological studies [58-61], i.e. protein/protein interactions [62-64], identification of targets for drugs [65-67], chemical reaction network [68-69], metabolic engineering[70-72] and *so on*. Thus, it is also interesting to capture the charge transfer network in a human-accessible, topological picture[73]. The topology analysis of the charge transfer network may inspire us to partition the QM/MM boundary, for which the charge transfer between the active site and the surrounding environment can be minimized *via* graph theory.

In this work, we propose a computational protocol for the construction and topology analysis of charge transfer network of any possible protein structure. This is derived from the single electron motion equation of bio-molecule systems under the tight binding approximation. Then, the issue of QM/MM boundary determination is converted into a complex network problem, which can be solved by graph theory. The charge transfer topology associated domain (ctTAD) or community structure is proposed as the optimal QM region. And we suggest the charge transfer network provides us an alternative tool for eliminating the QM/MM boundary charge transfer effects, which is computational efficient in a convenient and intuitive manner.

## 2. Theory and Computational Details

### 2.1 Derivation of Tight Binding Method for Bio-molecules

Here, the single electron motion equation for bio-molecule systems is derived according to the idea of tight-binding approximation. A brief introduction of the tight binding method for bio-molecules (bioTB) is mainly following the previous work of Liu and co-workers[74-76]. The transfer integral describes the ability to perform the charge transfer among neighbor sites, meanwhile, the on-site energy describes the ability to move or inject an electron from a specific site. The effective transfer integral can be given,

$$t^{eff}_{n,n+1} = \frac{t_{n,n+1} - \frac{1}{2}(\varepsilon_i + \varepsilon_j)s_{n,n+1}}{1 - s^2_{n,n+1}}$$

In the above, *s* is the orbital overlap integral between sites. This transformation shows minor effects on the on-site energy and therefore can be ignored in our code implementation. The parameters for the transfer integral in this model can be directly obtained from ab initio calculations, which can be directly applied to the electronic structure calculations of bio-molecules.

**2.2 The Community Structure and Charge Transfer Based Network**

The electronic Hamiltonian is not a tridiagonal matrix as the case of linear molecules, due to the complex stacking of amino acids in realistic proteins. To understand the irregular electronic Hamiltonian matrix, we attempt to construct the charge transfer knowledge graph or network in a more convenient and intuitive fashion. Here, without losing any generality, we will restrict our study to electron hole transfer coupling between HOMOs of amino acid combinations.

Electronic coupling elements as an important component for biological charge transfer can be derived from the above discussions. The abundance of high performance computing resources enables us to directly calculate the charge transfer couplings parameters for tens of thousands of molecular fragments, which represents possible occurrences in proteins.

In order to construct the protein charge transfer network, each residue is represented by a vertex in the graph, and the edge represents the strength of charge transfer coupling among residues. The charge transfer rate is proportional to the square of charge transfer coupling strength. [51, 54-55, 77-79] As we are primarily concerned with the network topology, we define a resolution for the network, for which the unsigned charge transfer coupling below a threshold value is assigned to be zero. For a specific graph, the edge can be only possible to be 0 or 1. The threshold of significant charge transfer coupling is set to be 0.01 eV in this work. The computational protocol is summarized in Scheme 1.

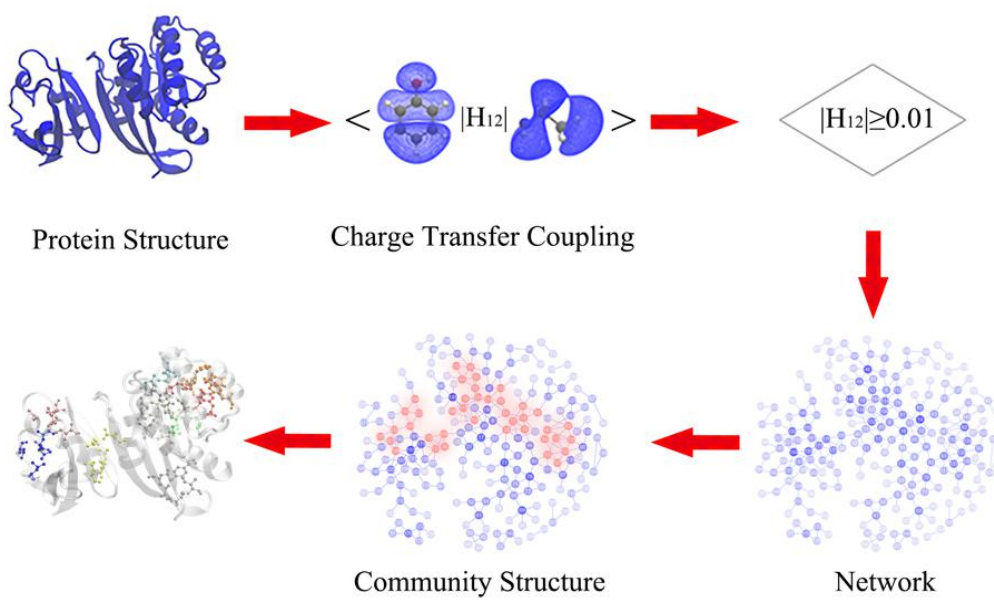

**Scheme 1**. The work flow for constructing the charge transfer network from charge transfer couplings. Once the charge transfer network is obtained, the network analysis and other statistics can be derived by a variety of means.

**2.3 Simulation Details**

As an initial evaluation, we randomly select two possible proteins from the PDB. One is the HIV-1 integrase core domain [80], and the other is the complex between the human H-Ras protein and the Ras-binding domain of C-Raf1 (Ras-Raf), which is central to the signal transduction cascade[81-82]. The molecular structure was minimized with subsequent 10 ns molecular dynamic simulation at 300K in the AmberTools package [83]. The final snapshot is taken for our subsequent analysis.

Then, we construct a data set of amino acid side-chain interactions for both proteins, and the procedure to extract each dimer complex has been described in the work of Thornton and co-workers.[84] Since the charge transfer couplings decay exponentially, we only take into account of the possible orientations of amino acid dimers within 10 Å of each other. The data set comprised 4702 and 7944 possible amino acid side-chain combinations for the protein monomer and dimer. The point of cutting covalent bond for the amino acid side-chain is saturated with hydrogen atoms (i.e., either the Cα or Cβ atom), and the missing hydrogens were added using the *tleap* module in AmberTools package. The data sets for the protein monomer and dimer are available at http://github.com/dulikai/bidiu.

All QM/MM simulations were carried out using Gaussian 09[85] for the QM portion and Amber[83] for the MM component. The QM region is modeled with HF and B3LYP calculations. The partial charges from Merz-Kollman scheme are used for evaluate the convergence of QM/MM calculations. The basis set sensitivity is determined by comparison with on two basis sets, namely 6-31G(d) and 6-311G(d,p) basis sets.

The site potential in Eq. 6 is obtained from the self-consistent filed calculations of the amino acid side-chain dimer. Both this potential and frontier orbitals of individual fragments were calculated at the HF/6-31G* and B3LYP/6-31G* levels. The monomer orbitals derived from HF or DFT calculations are non-orthogonal, for the orthogonal basis set is obtained from Lowdin's transformation. The computational protocol has been automated on a distributed computing cluster via a series of Python codes.

**Results**

Although previous studies[86-88] have revealed the relative abundance of various modes of amino acid contacts (van der Waals contacts, hydrogen bonds), relatively little is known about the qualitative charge transfer coupling terms of these noncovalent interactions. To evaluate the conformation ensemble influence, we calculate the charge transfer couplings for each possible amino acid side-chain combination in both proteins. The importance of each type of amino acids combinations is highlighted by a graph *via* the statistical representation of the charge transfer network. To facilitate our following discussions, the words "hot" and "cold" are used to describe the residues with larger and smaller degree of the node in the network.

In Figure 1, the graph is presented with the circular layout. Each vertex represents one type of amino acids. The edge represents the charge transfer ability between amino acids. The weight is assigned as the number of structures with charge transfer couplings larger than 0.01 eV in each kind of amino acid combinations. The width of the edge between the nodes is linearly corresponding to the edge weight, which is scaled by the total number of amino acid pairs in proteins. Thus, we could provide a global view of charge transfer couplings among twenty natural amino acids (totally 20×20 combinations).

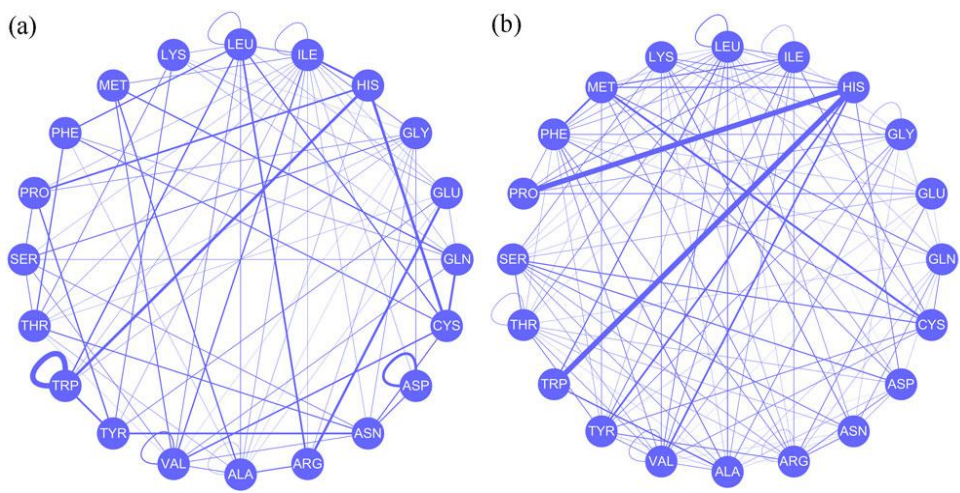

**Figure 1**. Knowledge graph of the charge transfer couplings by residue types at the resolution of 0.01 eV. (a) The HIV-1 integrase core domain. (b) The protein dimer (Ras-Raf). Note that, the weight is scaled by the total number of amino acid pairs in proteins.

The overall topology of charge transfer couplings distribution is strongly related to the physicochemical properties of amino acids. The comparison of the protein monomer and dimer indicates that each protein exhibits its specific charge transfer topological feature or fingerprint. The remarkable charge transfer couplings are mainly caused by a few specific types of amino acids, i.e. Trp, His, Pro, Asp and *so on*. In addition, the charge transfer coupling parameters are found to be very sensitive to the structural orientation of the amino acid side-chain combinations (Figure S1). Thus, the charge transfer effects for QM/MM boundary are dominated by these amino acids, although the contribution from other amino acids is not zero. Therefore, we may use this graph to rule out some impossible charge transfer routines.

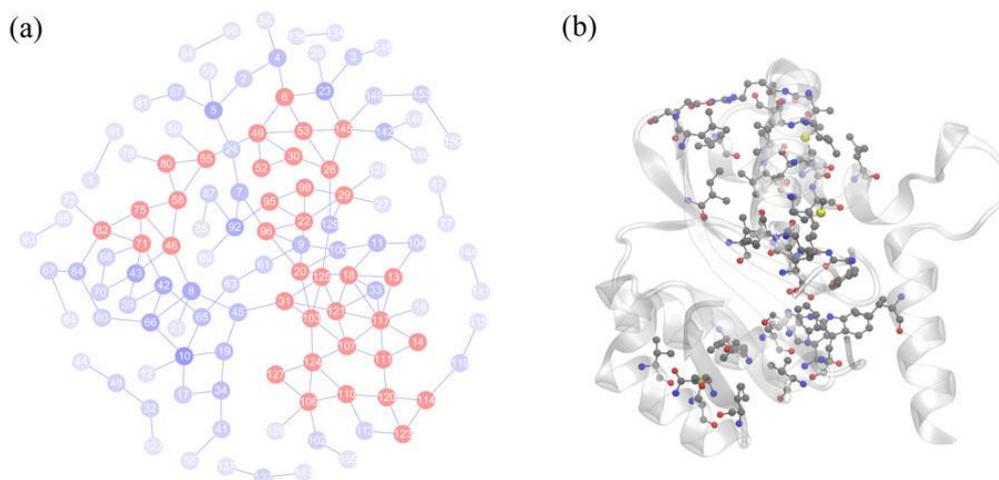

**Figure 2**. The charge transfer network by residues is given for the HIV-1 integrase core domain. (a) The charge transfer network at the resolution of 0.01 eV. (b) The node with high degree (red) is mapped to the three-dimensional protein structure. The red color refers to the hot residues with large node degrees, while the blue color means cold residue with very small node degrees.

Figure 2 presents the charge transfer network analysis of the HIV-1 integrase core domain. The CoSE layout[89] is applied to visualize the topological feature of un-directed graphs, which highlights the most important nodes (hot residues) and their surrounding nodes in an intuitive way. Note that, the charge transfer network is also given for the protein monomer and dimer with a resolution of 0.02 eV in Figure S2.

In Figure 2a, the hot residues with deep red color forms a reduced version of the protein charge transfer core. And the remainder of the cold residues with light red color does not substantially affect charge transfer when included. This also indicates the charge transfer importance of each residues. It is clear that the hot residues are not continuous in a specific secondary structure. The cutting of the suitable residues in the protein sequences could avoid the significant charge transfer effect between two neighboring amino acids. Figure 2b maps the node with high degree distribution in the charge transfer network on to the three-dimensional protein structure. The hot residues with significant charge transfer contributions in the protein structure are rendered in red color. This illustrates non-local topology properties of the charge transfer in proteins, which may be critical to understand the underlying protein charge transfer problems.

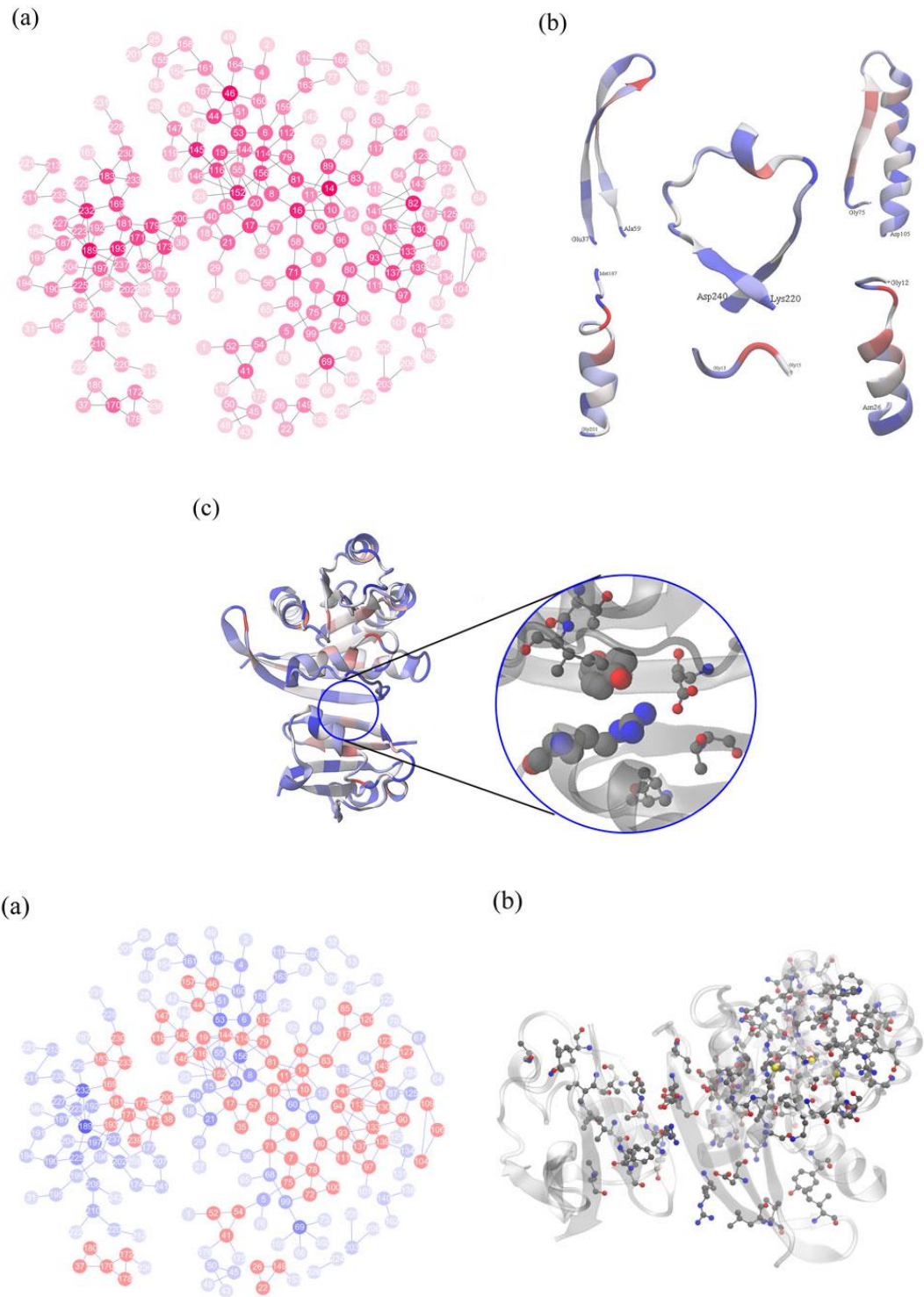

**Figure 3**. The charge transfer network by residues is given for the protein dimer (Ras-Raf). (a) The charge transfer network at the resolution of 0.01 eV. (b) The node degree distribution is mapped to the three-dimensional protein structure. The red color refers to the hot residues with large node degrees, while the blue color means cold residue with very small node degrees.

The charge transfer network analysis is also performed for the protein dimer (Ras-Raf) in Figure 3. As shown in Figure 3a, the charge transfer topology properties for this protein dimer is relatively different from that of the protein monomer in Figure 2. Figure 3a highlights the distinct boundary of the charge transfer network between two sub-units in the three-dimensional protein structure. Although the charge transfer interactions are significantly independent to each other, the hot residue at the protein-protein interface is indeed observed. In Figure 3c, the three-dimensional structure suggests the hot residues prefer to undergo charge transfer among distinct secondary structures. In addition, the hot residues are usually spaced and insulated by cold residues in the same secondary structure. This may also be the reason why we can successfully describe the QM/MM boundary across the amino acid sequences without remarkable charge transfer effects. The identification of charge transfer regions may provide us a possible solution to minimize the QM/MM boundary charge transfer effects.

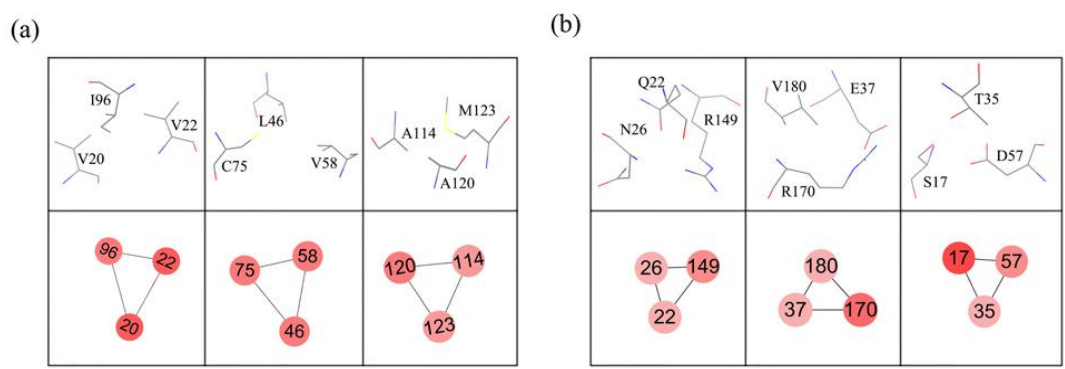

**Figure 4**. Typical sub-graphs and their corresponding molecular geometries in the charge transfer network of the protein monomer (a) and dimer (b).

Qualitatively, these hot residue groups with deep red color may be referred as somewhat "charge transfer topology associated domain" (ctTAD). As shown in Figure 4, we could easily identify a few typical motifs in the charge transfer network. The charge transfer network can be split into different regions near the less or weak connected nodes. And we suggested one could cut as less as possible edges to minimize the QM/MM boundary charge transfer effects. This can be easily achieved by the visual inspection of the network connections or community detection algorithms in graph theory.

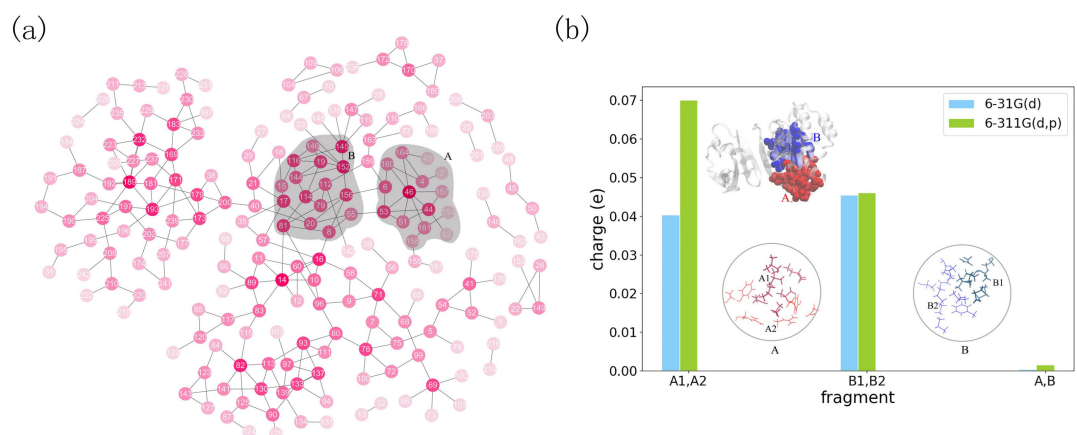

**Figure 5**. (a) The typical charge transfer topology associated domain (ctTAD) in the charge transfer network analysis of the Ras-Raf protein dimer at the resolution of 0.01 eV. (b) Atomic partial charge difference analysis between various selections of QM region in QM/MM calculations.

Atomic charges are a key concept to give more insight into the electronic structure and chemical reactivity. It has been reported that the convergence of atomic partial charges are very slow in a few recent QM/MM calculations。[18, 20] For demonstrative purposes, we try to use the ctTAD to minimize the partial charges variations across the QM/MM boundary.

Figure 5a illustrates two possible ctTADs (region A, B) in the charge transfer network. Briefly, partial charge population was analyzed for region A and B, and randomly separated A1, A2 and B1, B2. As shown in Figure 5b, the destruction of the ctTAD introduces significant charge transfer between A1 and A2 region, and the connections between B1 and B2 also introduce a remarkable charge transfer interactions. The partial charges variation between region A and B is significantly minor because they belong to different ctTADs. Note that, the larger basis set may cause more serious charge transfer problems in QM/MM calculations. In summary, the partial charge differences can easily produce small (ca. 0.01 – 0.02 e) atomic charges differences if the ctTAD is used as a criterion for QM/MM boundary partitions.

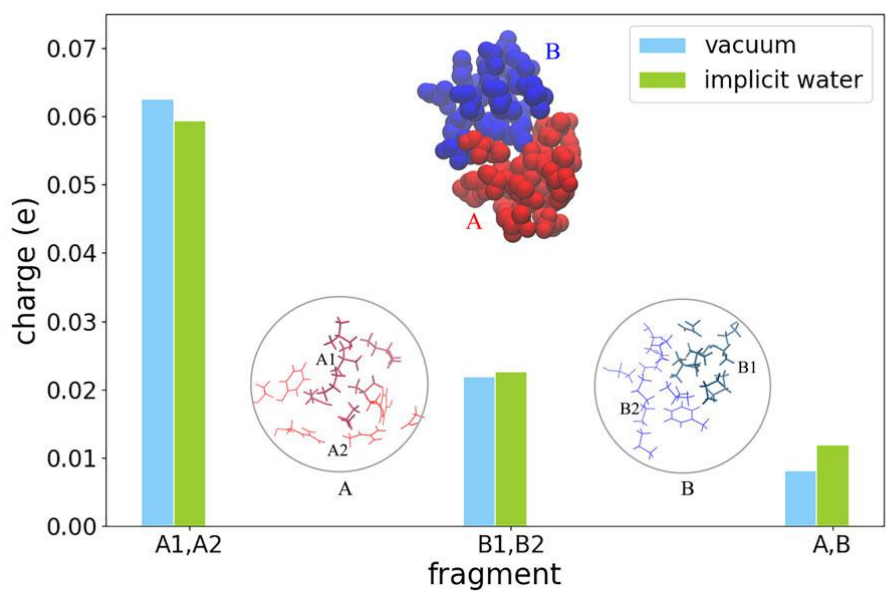

**Figure 6**. Atomic partial charge different analysis of the charge transfer between various residues selections in the quantum chemistry cluster model. Insets are the corresponding residues in the same ctTADs.

The quantum chemical cluster approach is an alternative choice for modeling enzyme active sites.[90-93] We believe the ctTAD from network analysis is also is of great benefit to chemical cluster model. Figure 6 shows the performance of the ctTADs in the quantum chemistry cluster model. It is clear that the ctTAD analysis also alleviates charge transfer effects for the residues selection of chemical cluster models. Therefore, the network representation of charge transfer in realistic proteins provides us a useful way to understand the topology of charge transfer effects, which may be helpful to minimize the QM/MM boundary, as well as the chemical cluster model.

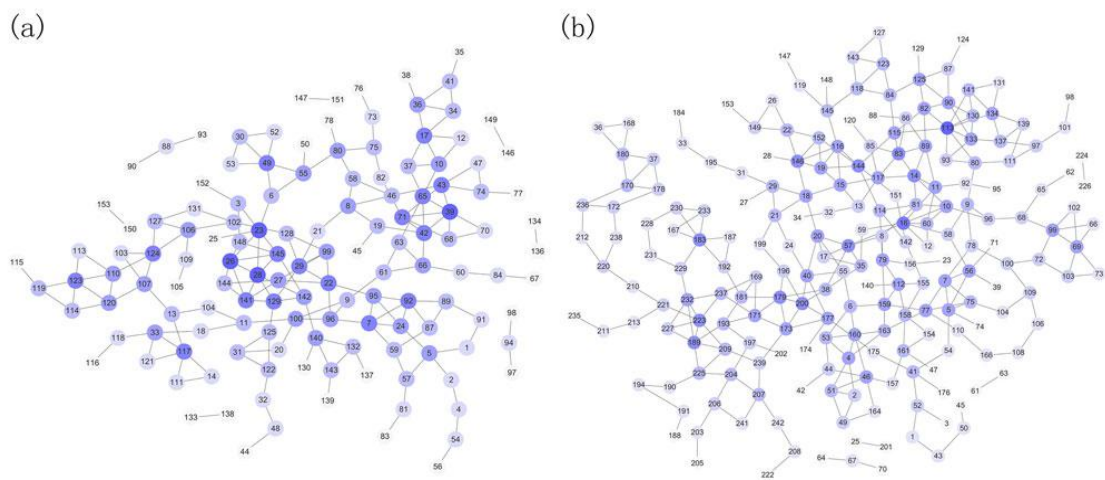

**Figure 7**. The network derived from amino acid radical distances for the protein monomer (a) and dimer (b) at the resolution of 6.0 Å.

The existence of ctTADs inspires us to search for its structural counterparts. For example, can we find an optimal structural topology associated domain (sTAD) on the basis of radical distance cutoff? Traditionally, the determination of QM/MM region is achieved by the radial distance cutoffs from a central point of interest in a QM/MM calculation. If one can find such a sTAD with similar performance as the ctTAD, it is would be very helpful to simply use the sTAD to minimize the charge transfer effects across the QM/MM boundary.

In general, the radical distance among residues can also be represented by a graph (Figure 7). Thus, we define a resolution for the distance graph, for which the radical distances between residues larger than a threshold value is assigned to be zero. Then, we tried to tune the distance network at different resolution. The resolution of 6.0 Å seems to provide similar topological parameters as the charge transfer network (Table S1). And we also provide the distance network with a resolution of 5.0 Å in Figure S3. For our subsequent discussions, the distance network is given for the protein monomer and dimer with a resolution of 6.0 Å.

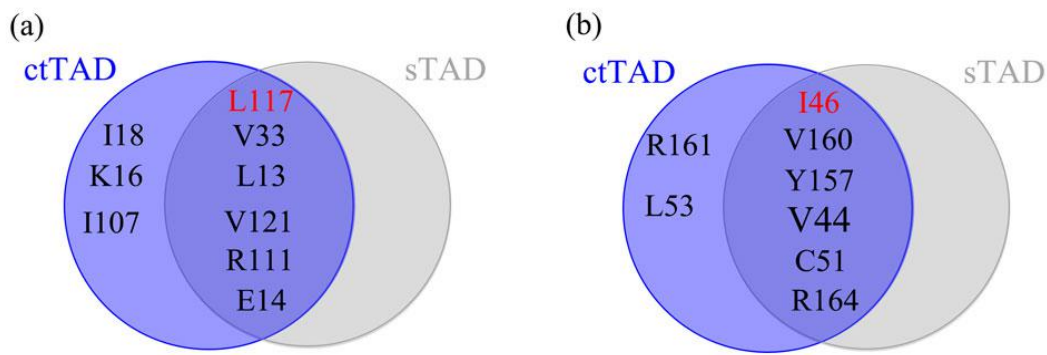

**Figure 8**. Venn diagram is applied to compare the residues sets of ctTAD and sTAD, for the protein monomer (a) and dimer (b). The ctTAD is given at the resolution of 0.01 eV, and the sTAD is optimized at the resolution of 6.0 Å. The residues with red color indicate the central residue in the topology associated domain.

The sTAD could also be obtained by analyzing the node connections or community detection algorithms. However, we cannot find the same sets of residues for sTAD with ctTAD as shown in Figure 8. A few extra amino acids are detected by the ctTADs, which should be included in the QM regions. Some of these extra residues with larger charge transfer contributions are expected, such as the polar arginine and lysine (R161, K16). Meanwhile, the detection of the residues I18, I107 and L53 is surprising since they are both non-polar and not the closest to the center of the central residues. We noted that similar observations have also been reported in the determination of QM/MM boundary by the CSA and FSA methods.[26] Thus, the connected node in the charge transfer network is not necessary to the closest, and the identified residues are not always those that would be typically selected using criteria such as chemical intuition or proximity.

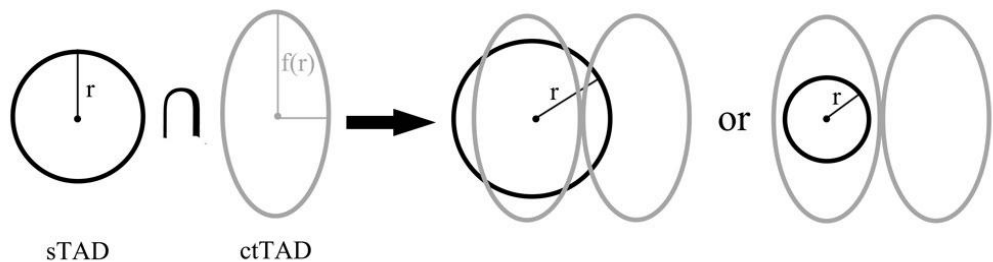

**Figure 9**. We use the sphere to describe the physical nature of the sTAD scheme, and the ellipsoid can be used to describe the physical nature of the ctTAD scheme. The sTAD scheme would often incorporate more or less residues in contrast to the ctTAD scheme.

The ctTAD can explicitly describe the intrinsic anisotropic feature of the charge transfer effects among amino acids. This provides an intuitive perspective for the electronic structure of the protein complex. For example, the increase of the resolution for sTAD, i.e. $\geq 6.0$ Å would enhance the charge transfer across QM/MM boundary. This is because the larger sTAD would be across two ctTADs, and the destruction of ctTADs is one of the significant sources of charge transfer errors. In summary, the radical distance cutoff only partially reflects the charge transfer network in realistic proteins. And the QM regions selected by radial distance alone may be failed to detect some residues that are important for the charge transfer effects of the QM/MM boundary.

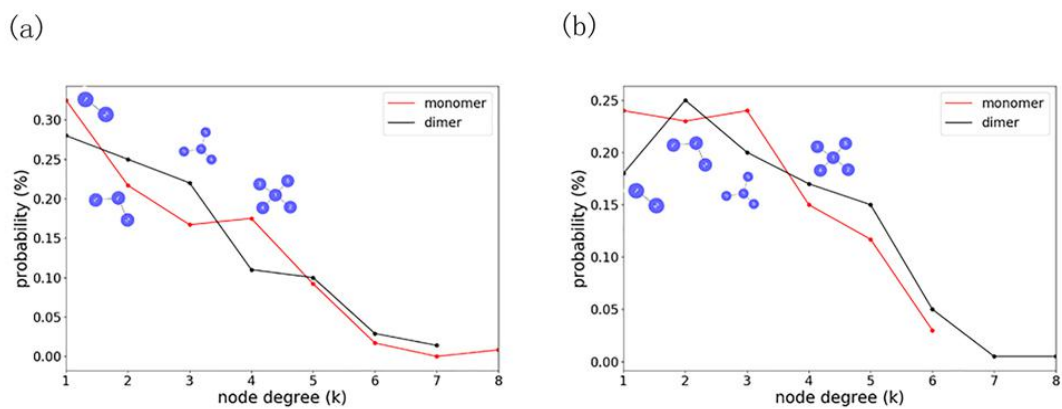

**Figure 10**. The degree distribution for the charge transfer networks (a) and distance networks (b). Both the protein monomer and dimer are analyzed.

The nodes degree distribution is very important in studying realistic networks. Figure 10 shows the degree distribution for the charge transfer network and distance network. The charge transfer network as many realistic network in biology or social science, are neither regular network nor random network. Typically, the degree distribution of charge transfer network follows a power law, at least asymptotically. The fitted exponent is in the range of 1.0~1.5. And we suggest the charge transfer network can be generally regarded as a scale-free network[94-95], for which the most critical residues act as network hubs for the partition of the QM/MM boundary. Meanwhile, the degree distribution of distance network remarkably biases from the scale-free network. It seems the distance network is similar as a random graph, although the node degree distribution is highly left-skewed. Thus, the topology of the network from structure and charge transfer properties is different in nature. This may explain the inconsistent between the sTAD and ctTAD in the QM/MM boundary partition.

Inspired by the concept of chemical bonds, we may also refer the connections in the charge transfer network as topology associated charge transfer bonds. And the residues with more connections in the charge transfer network can be regarded as topology de-localized charge transfer bonds, which should be avoids to be separated across the QM/MM boundary in most cases. Meanwhile, the residues with only one or two connections are topology localized charge transfer bonds, which are encouraged to be separated across the QM/MM boundary. Because the scale-free nature of the charge transfer network, topology associated charge transfer bonds mainly have less than 4 connections (more than 90%). And the decay of the node degree distribution of the charge transfer network is much significant than that of the distance network. Thus, one can easily find the suitable QM/MM boundary, which satisfies the criterion of cutting less node connections. This may be also the reason why the QM/MM method is still robustness in various simulations with tens of atoms in the QM regions. Further studies may be required to understand this interesting topology feature.

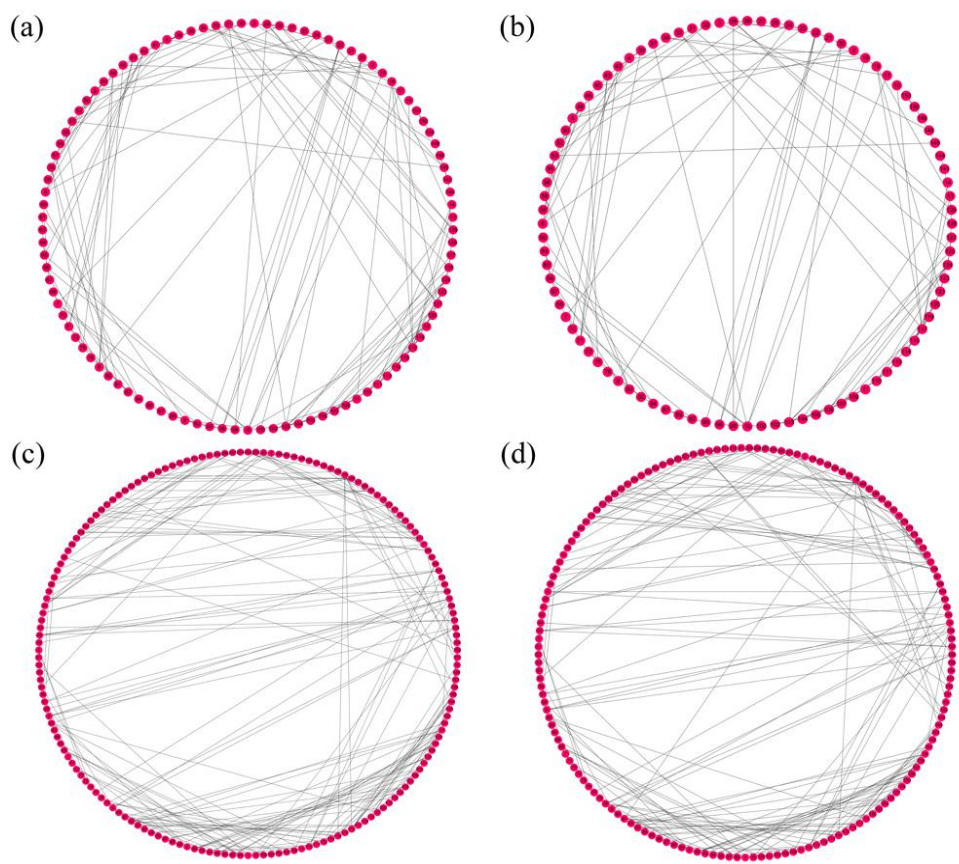

**Figure 11**. The charge transfer network among residues is given at the resolution of 0.01 eV. (a) The protein monomer at HF level; (b) the protein monomer at B3LYP level; (c) the protein dimer at HF level; (d) the protein dimer at B3LYP level. The graph representations of protein charge transfer network are given with circular layout.

Finally, the topology dependent on electronic structure methods for the charge transfer network is evaluated at HF and B3LYP level. In Figure 11, the circular layout is used to represent alternative view of the protein charge transfer network. The advantage of a circular layout in the biological applications is its neutrality, because none of the nodes (residues) is given a privileged position by placing all vertices at equal distances from each other as a regular polygon. Thus, the circular layout views in protein systems may provide useful insights on the topology relationship among different methods. The network topology parameters are summarized in Table S2. And we find that the possible electronic method changes do not significantly influence the electronic properties in proteins. Therefore, the topology of the charge transfer network is stable to the electronic structure method variations.

**Discussions**

The complex network analysis is very common in the realm of biological science. Inspired by this issue, we use the language of graph theory to evaluate and visualize the importance of charge transfer between QM/MM boundaries, which avoids the complicated matrix manipulations in the electronic structure calculations. This scheme is strictly derived from the Schrödinger equation under the tight-binding approximation, and the parameters are obtained directly from ab initio calculations, with clear physical meanings.

The graph theory reveals the ctTAD as an instructive and efficient indicator to partition the QM and MM region. The detection of ctTAD is very easy by visual inspection or community detection algorithm in graph theory. In the validation strategy, we prove the integrity of ctTAD is very important to minimize the atomic partial charges across the QM/MM boundary. The typical ctTAD usually comprises ~100-200 atoms, less than 10 residues, which could be possible for most available quantum chemistry package.

Note that, the integrity of the ctTAD should be retained in QM/MM boundary partitions. The widely used radial distance cutoffs, namely sTAD in this work, cannot completely reproduce the same residues sets as the ctTAD, without disrupting other ctTADs. The destruction of ctTADs by radial distance may partially cause the slow convergence problem, and large fluctuation in the calculations of molecular properties. The inconsistent between ctTAD and sTAD suggests that the QM regions are not always successfully chosen by the radical distance scheme. This is expected, because the charge transfer interaction is not simply a linear function of molecular structures. Therefore, the QM regions determination scheme based on radical distance is not sufficient. And the electronic structure based descriptors, such as the charge transfer coupling, partial charge or Fukui function should be considered in order to systematically determine QM/MM regions.[16, 21-26]

The scale-free property of charge transfer network strongly correlates with the network's robustness to failure. In this work, the high degree nodes are rare and can be thought to serve specific purposes in the networks. This network model can be easily implemented in a graphical user interface (GUI), for which one can use the mouse and touch screen to select optimized QM/MM regions with minimized inter-region charge transfer interactions. Each ctTAD as a community structure in the network can be viewed as a local excitation region, which should be treated as an entity. Although the charge transfer coupling is not the sole factors to determine the charge transfer effects, one can use it to roughly rule out most impossible routines.

A further consideration is the computational cost. Because each calculation in this work consists of only pairwise residues, it becomes straightforward to rapidly scan through all residues in the protein in a highly parallelized fashion. The computational protocol for obtaining ctTAD can be accomplished within only a few minutes.

**Conclusions**

The reasonable validation of charge transfer between the QM and MM region for the hybrid QM/MM method is important for partition the core active site and the surrounding protein environment. In this work, we derive the single electron motion equation for the bio-molecules on the basis of tight-binding approximation. Beyond directly solving the electronic structure equation, we suggest to use the charge transfer knowledge graph obtained from the bioTB model to quickly judge possible charge transfer preference in realistic proteins.

In summary, the concept of ctTAD provides a robust framework to systematically design and interpret the charge transfer effects on the QM/MM boundary, which are expected to provide broad utility for improving the robustness of multiscale modeling efforts. The ctTAD also offers us an alternative approach for efficiently identifying the critical residue or groups of residues to minimize the QM/MM boundary, which avoids wasting computational resources. Further work is also going on to verify the ability of ctTAD in the catalytic reaction mechanism and photo-induced process of condense phases.

**Acknowledgments**

The authors thank the support by the National Natural Science Foundation of China (Nos. 21503249). L. Du also thanks the positive discussions with Beijing PARATERA Tech Co., Ltd.

**Supporting information**

The structural dependence of charge transfer couplings, network and topological parameters are given in are given in the supporting information.

TOC Graphic

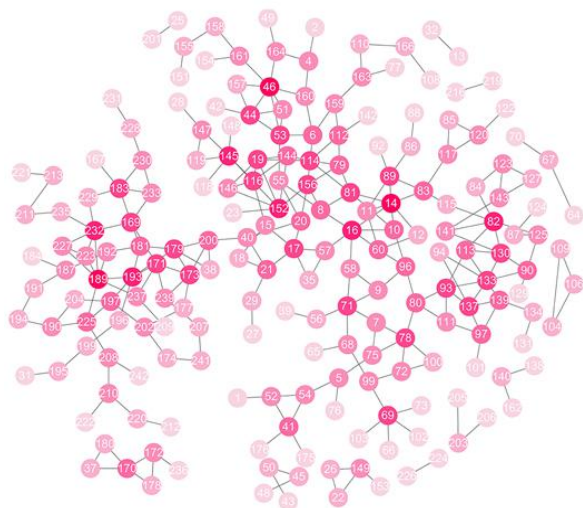
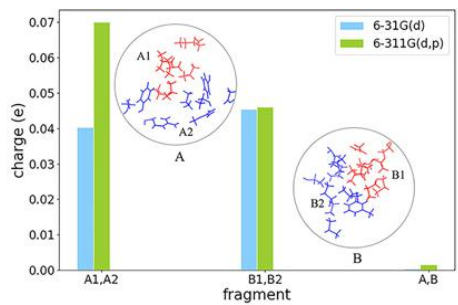
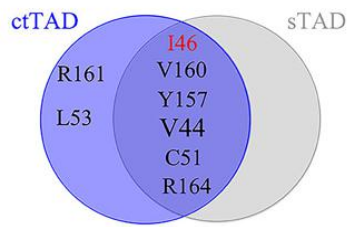

# Supporting Information

**Electronic Structure Topology Associated Domain is Useful to Minimize the Uncertainty of QM/MM Boundary Charge Transfer Effects**

Jiajun Yang[1], Fang Liu[1], Dongju Zhang[2], Likai Du[1]*

[1]Hubei Key Laboratory of Agricultural Bioinformatics, College of Informatics, Huazhong Agricultural University, Wuhan, 430070, P. R. China

[2]Institute of Theoretical Chemistry, Shandong University, Jinan, 250100, P. R. China

*To whom correspondence should be addressed.*

*Likai Du: dulikai@mail.hzau.edu.cn*

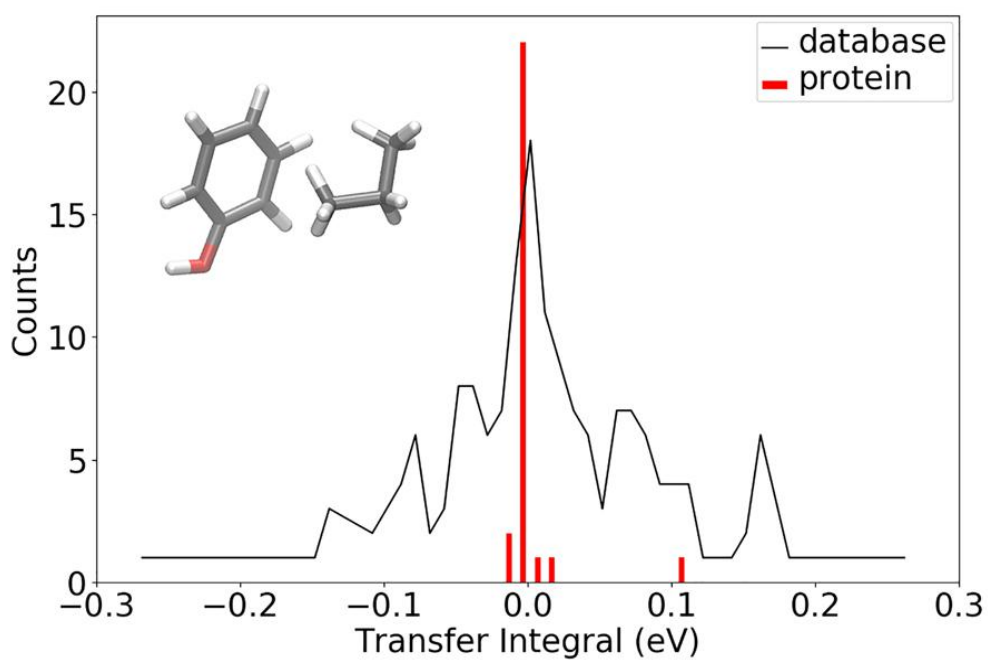

**Figure S1**. The distribution of charge transfer couplings (eV) for the Tyr/Ile pair are shown, whereas Tyr is the center fragment. The red line refers to the charge transfer couplings for the studied proteins in this work. For comparison, the black curve represents the charge transfer couplings from possible structures of Tyr/Ile pairs derived from more than 2000 proteins structures, which is also available from http://github.com/dulikai/bidiu.

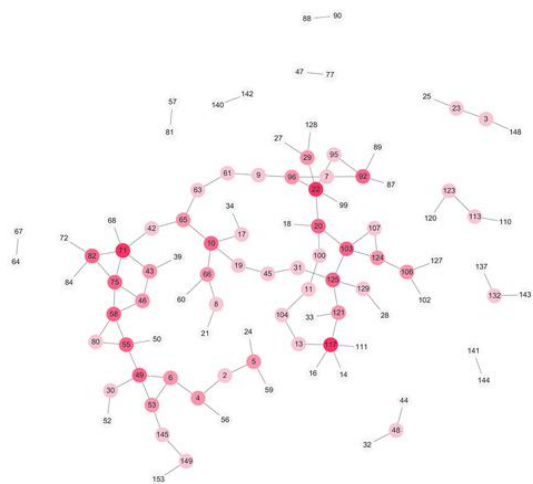 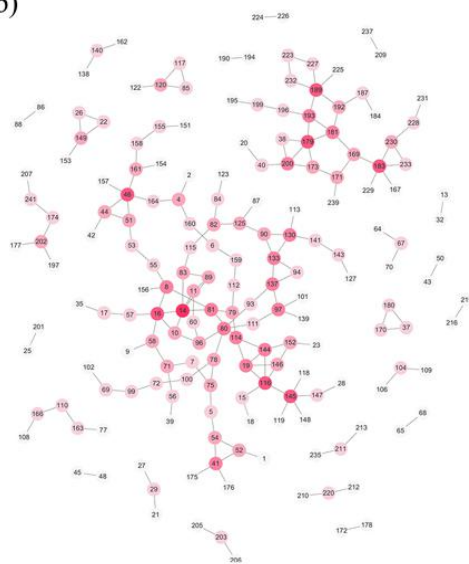

**Figure S2**. The charge transfer network for the protein monomer (a) and dimer (b) at the resolution of 0.02 eV.

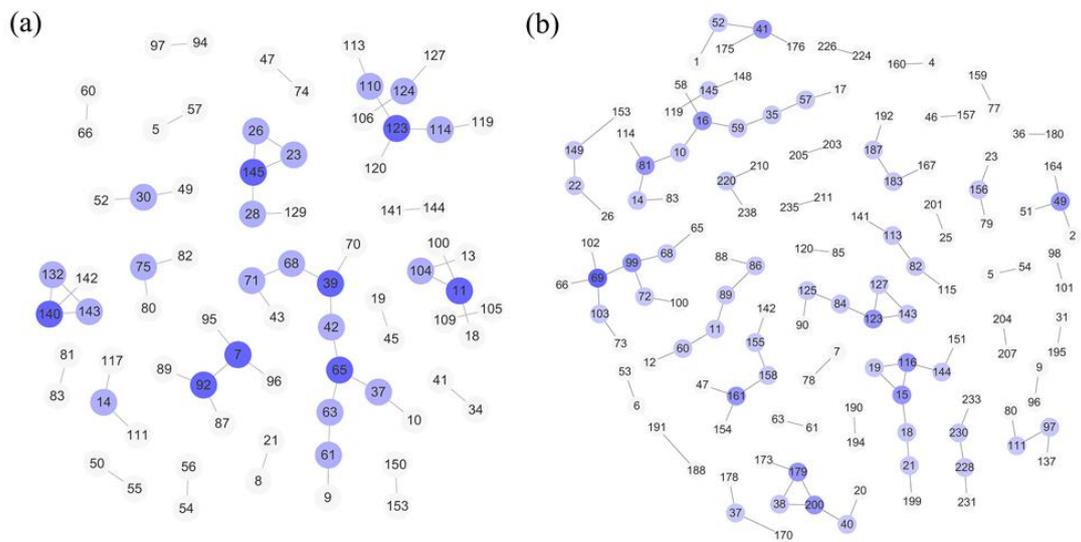

**Figure S3**. The network derived from amino acid radical distances for the protein monomer (a) and dimer (b) at the resolution of 5.0 Å.

**Table S1**. The topology parameters of the distance network and charge transfer network at the resolution of 6.0 Å. The residues with zero connections are omitted.

|  | Ras-Raf dimer | | HIV-1 integrase core domain | |
|---|---|---|---|---|
|  | ctTAD | sTAD | ctTAD | sTAD |
| Number of nodes | 204 | 212 | 120 | 137 |
| Network density | 0.013 | 0.014 | 0.022 | 0.020 |
| Network centralization | 0.022 | 0.024 | 0.046 | 0.024 |
| Clustering coefficient | 0.177 | 0.192 | 0.10625 | 0.229 |
| Avg. Number of neighbors | 2.657 | 3.047 | 2.583 | 2.759 |
| Network heterogeneity | 0.566 | 0.503 | 0.514 | 0.580 |

**Table S2**. The topology parameters of the charge transfer network at the resolution of 0.01 eV with HF and B3LYP methods. The residues with zero connections are omitted.

|  | Ras-Raf dimer | | HIV-1 integrase core domain | |
| --- | --- | --- | --- | --- |
|  | HF | B3LYP | HF | B3LYP |
| Number of nodes | 170 | 161 | 101 | 92 |
| Network density | 0.017 | 0.017 | 0.029 | 0.027 |
| Network centralization | 0.025 | 0.033 | 0.053 | 0.062 |
| Clustering coefficient | 0.159 | 0.137 | 0.126 | 0.159 |
| Avg. Number of neighbors | 2.871 | 2.720 | 2.851 | 2.478 |
| Network heterogeneity | 0.532 | 0.550 | 0.518 | 0.553 |